**Acoustic superscatterers for passive suppression of cylindrical source radiation in the forward direction**

Vineeth P. Ramachandran[a][b], Prabhu Rajagopal[b]#

[a]*DRDO Young Scientists' Laboratory for Smart Materials, Hyderabad, India-500058*

[b]*Centre for Non-destructive Evaluation and Department of Mechanical Engineering, Indian Institute of Technology- Madras, Chennai, India-600036*

#Corresponding author: prajagopal@iitm.ac.in

**Abstract:**

Superscatterers are known to expand the rigid boundary of an object thereby enhancing the scattering cross section of the object. The design philosophy of the acoustic superscatterer is based on the 'partially-resonant system' in which a coating material, i.e.; double-negative metamaterial, complementary to the host medium is provided on top of the rigid object. When the source lies within the enhanced boundary of the complementary media, the interaction between the radiated wave-front and the enhanced virtual boundary gets stronger and results in suppression of the total forward radiated sound at far-field. An analytical framework is shown in this paper on suppression of the total acoustic pressure at far field in the forward direction when a cylindrical source, both monopole and dipole, lies within the virtual rigid boundary of the superscatterer. Total extinction cross section of the scatterer as a function of the distance between the source and scatterer is derived to substantiate the earlier results. Additionally, the effect of the area of cross-section of the cylindrical source on the total forward radiated sound pressure at far-field is discussed. Finally, the effectiveness of the analytical results is verified numerically and the prospects for practical applications are discussed.

1.  **Introduction:**

Suppression of noise is vital to many branches of science and engineering, and various active and passive methods are reported in literature and realised in practice for this purpose. Active noise control (ANC) systems use a loudspeaker or actuator to broadcast a noise-cancelling signal at the desired location so that noise attenuation can be achieved remotely. Typically, ANC systems require a reference sensor, an actuator, and an error sensor. Noise reduction at the patient's ear during magnetic resonance imaging without any headphone is reportedly achieved by ANC systems such as non-ferromagnetic equipment by cancellation of the sound at the patient's ears [1-2]. Xiao *et al.* [3] overcame the difficulty of placing an error hydrophone in a person's ear by suitably mounting a 'virtual ANC hydrophone' in the form of a laser pick-up and could achieve a sound attenuation of 10 dB in 0.5-6kHz frequency band. Trent *et al.* [4] explored remote delivery of a noise-cancelling signal to a desired location using the time reversal (TR) inverse filter technique with a $180^0$ phase shift since TR is capable of remotely focusing sound energy to a point in space. Durant *et al.* [5] presented a study on ANC in duct with a Harmonic Acoustic Pneumatic Source (HAPS), designed to generate a controlled harmonic anti-noise from a pneumatic source. Here, in contrast to one signal driving a loud speaker of large band-width, two signals of very low bandwidth each for amplitude and frequency of anti-noise drive were proposed. However, the utility of such ANC mentioned above is limited by poor efficiency, low compactness and the operating conditions (high temperature, airflow velocity for example).

Passive noise control (PNC) is achieved by insulating the source typically with porous absorbing materials or by the provision of mufflers [6] or resonators. In PNC, the acoustic causality constrains [7] the thickness and density of the absorbing material, thus leading to limited performance at lower frequencies. Good sound absorption warrants denser and thicker acoustic materials. Most porous materials commonly used for absorption have poor sound insulation. Jena *et al.* [8] explored the possibility of periodic rigid scatterers placed inside porous materials to improve their transmission loss (TL) with the Bragg diffraction. They argued that the TL of the material is increased significantly due to scattering between the scatterers within the porous medium. Sonic crystals or periodic scatterers can significantly increase sound attenuation within Bragg resonance zone, depending upon the number of rows of periodic arrangement [9-10]. Hybrid configurations such as scatterers with porous cores [11-12] or scatterers with acoustic double panels [13] have also been proposed to increase the TL.

In recent years, several researchers have considered metamaterial-based structures, for achieving effective damping of wave energy [14,15]. Sound absorption in Helmholtz metamaterial resonators [16-18] occurs in the viscous and thermal boundary layers formed at the walls of the resonator. Duan *et al.* [19] proposed a deep subwavelength (thickness is only 1/30[th] of the working wavelength) acoustic multi-layer Helmholtz resonance metamaterial, which can achieve multiple absorption peaks in a frequency band. Acoustic impedance regulation of a neck embedded Helmholtz resonator, reported by Duan *et al.* [20], could achieve 99.9% sound absorption at 158 Hz across a deep subwavelength thickness (1/42 of wavelength). Boccaccio *et al.* [21] proposed a deep subwavelength (total thickness less than 1/28 wavelength) hybrid parallel-arranged microperforated panel and Archimedean-inspired spiral (AIS) absorber to obtain broadband sound absorption above 60% at low frequency (400–2000 Hz). There has been concerted efforts to suppress the elastic wave propagation with the help of metamaterial waveguides by creating bandgap wherein higher harmonics of the wave is hindered [22-23]. Transformational optics led the path towards metamaterials to produce spatial variation such that electromagnetic waves can be redirected [24]. Since equations for acoustic waves are also form-invariant to coordinate transforms, acoustic metamaterials are also realisable using the coordinate transformation approach [25]. Our group [26-27] has recently used this approach for manipulation of a plane wavefront into a cylindrical wavefront in a metamaterial plate made of gradient refractive index phononic crystals.

'Superscatterers' are a class of metamaterials shown to enhance the electromagnetic wave scattering cross section of an object, for example a perfect electrical conductor cylinder (PEC), so that it looks bigger than the actual size of the object [28]. Such superscatterers, also called 'partially-resonant systems' [29], are theoretically realisable by coating with a negative refractive material shell complementary to the host medium, on a PEC. An acoustic corollary of the electromagnetic superscatterer has been shown to exist theoretically [30]. A recent numerical simulation study reported [31] that acoustic superscatterers can also be employed for omni-directional suppression of acoustic radiation when the radiating source lies within the expanded boundary of the scatterer.

Here for the first time, we discuss a complete analytical model of suppression of radiated sound at far-field in the forward direction from a monopolar and dipolar oscillating cylinder using an acoustic superscatterer. This paper is organised as follows.

## 2. Theory:

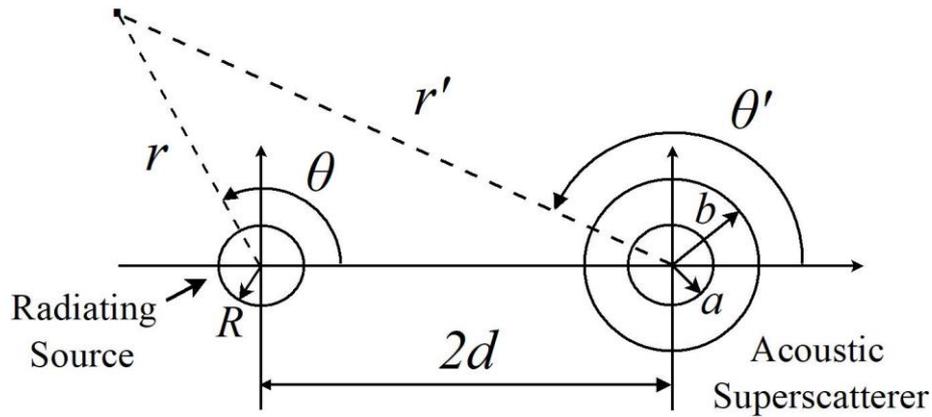

**Fig 1:** Schematic illustration of the geometry of the problem under study

Consider a cylinder of mean radius $R$ undergoing axisymmetric harmonic vibration in a host medium with bulk modulus $\lambda_0$ and density $\rho_0$. The radiated pressure field from this active source at any point $(r,\theta)$ at instance $t$ is,

$$P_{rad}(r,\theta,t) = p_0 e^{-i\omega t} \sum_{n=-\infty}^{\infty} C_n H_n^{(1)}(k_0 r) e^{in\theta} \tag{1}$$

where $p_0$ is the pressure amplitude, $\omega$ is the radial frequency of vibration, $H_n^{(1)}(.)$ is the cylindrical Hankel function of first kind of order $n$, and $C_n$ is the undetermined expansion coefficient that will be determined from the suitable boundary and continuity conditions.

Consider a rigid cylinder ($r' \leq a$) coated with a shell ($a \leq r' \leq b$), with its centre located at a distance $2d$ from the centre of the active source. The acoustic wave equation in the cylindrical co-ordinate system within the shell is,

$$\frac{\lambda(r')}{r'} \frac{\partial}{\partial r'}\left[\frac{r'}{\rho_{r'}} \frac{\partial p}{\partial r'}\right] + \frac{\lambda(r')}{r'^2 \rho_{\theta'}} \frac{\partial^2 p}{\partial \theta'^2} + \omega^2 p = 0 \tag{2}$$

where $p(r',\theta',z,t)$ is the acoustic pressure, $\rho_{r'}$ and $\rho_{\theta'}$ are the radial and tangential components respectively of density tensor relative to $\rho_0$, and $\lambda(.)$ is the radially varying bulk modulus relative to $\lambda_0$.

By introducing coordinate transformation from $(r', \theta', z)$ to $(f(r'), \theta', z)$ while maintaining $p(r', \theta', z) = P(f(r'), \theta', z)$, one can easily rewrite Eqn. (2) as,

$$\frac{\lambda(r')f'(r')}{r'} \frac{\partial}{\partial f}\left[\frac{r'f'(r')}{\rho_{r'}} \frac{\partial P}{\partial f}\right] + \frac{\lambda(r')}{r'^2 \rho_{\theta'}} \frac{\partial^2 P}{\partial \theta'^2} + \omega^2 P = 0 \qquad (3)$$

where $f'(r') = \dfrac{\partial f(r')}{\partial r'}$

In order to preserve the form invariance of wave equation between the original and the new coordinates, the values of $\rho_{r'}, \rho_{\theta'}$ and $\lambda(r')$ can be taken as,

$$\rho_{r'} = \frac{r'}{f}\frac{\partial f}{\partial r'}, \quad \rho_{\theta'} = \frac{f}{r'}\frac{1}{\left(\partial f/\partial r'\right)}, \quad \lambda = \frac{r'}{f}\frac{1}{\left(\partial f/\partial r'\right)} \qquad (4)$$

The solution for Eqn. (3) is sought by using the separation of variables method, neglecting the $z$-directional dependence of acoustic pressure since the acoustic wave propagation is perpendicular to the cylinder axis, resulting in the pressure within the shell as,

$$p(r', \theta', t) = p_0 e^{-i\omega t} \sum_{n=-\infty}^{\infty} \left[E_n H_n^{(1)}\left(k_0 f(r')\right) + F_n J_n\left(k_0 f(r')\right)\right] e^{in\theta'}, a < r' < b \qquad (5)$$

where $E_n$ and $F_n$ are the undetermined expansion coefficients that will be determined from the suitable boundary conditions, $J_n(.)$ is the cylindrical Bessel function of first kind of order $n$. The additional term $F_n J_n\left(k_0 f(r')\right)$ in Eqn. (5) is provided so that $E_n$ is not vanishing at the rigid boundary ($r' = a$).

At distances $r' > b$, the scattered pressure from rigid cylinder coated with shell is,

$$p_{scat}(r', \theta', t) = p_0 e^{-i\omega t} \sum_{n=-\infty}^{\infty} D_n H_n^{(1)}\left(k_0 r'\right) e^{in\theta'}, r' > b \qquad (6)$$

where $D_n$ is the undetermined expansion coefficient that will be determined from the suitable boundary conditions.

In order to satisfy the condition for a partially-resonant system [5], i.e.; $\lambda(r'=b)\lambda_0 + \lambda_0 = 0$, the transformation function for the coating is selected as $f(r') = \dfrac{b^2}{r'}$. Therefore, the coating material is a complementary media to the host medium and we have,

$$\rho_{r'} = -1, \; \rho_{\theta'} = -1, \; \lambda = -\dfrac{r'^4}{b^4} \tag{7}$$

Using Eqn. (7), the bulk modulus of the coating material at the outer periphery is $\lambda(r'=b) = -1$. Hence a 'partially-resonant system' is established with coating material, a double-negative-metamaterial (both the density and bulk modulus indices are negative), and the host medium having complementary material indices. Now the rigid cylinder of radius $r'=a$ is magnified to radius $f(r'=a) = \dfrac{b^2}{a}$ in the transformed coordinates while maintaining $r'=b=f(r'=b)$.

Using the translational addition theorem [32,33,34] as shown in Fig. 2 below, Eqn. (1) and Eqn. (6) can be referred to $(r',\theta')$ and $(r,\theta)$ coordinate systems respectively.

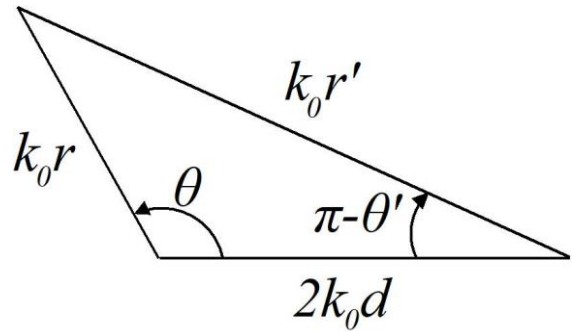

**Fig 2**: Geometry requirement for the application of translational addition theorem [34]

Eqn. (1) is referred to $(r',\theta')$ coordinate system as,

$$P_{rad}(r',\theta',t) = p_0 e^{-i\omega t} \sum_{n=-\infty}^{\infty} \left( \sum_{m=-\infty}^{\infty} C_m H^{(1)}_{m-n}(2k_0 d) \right) J_n(k_0 r') e^{in\theta'} \tag{8}$$

Eqn. (6) is referred to $(r,\theta)$ coordinate system as,

$$p_{scat}(r,\theta,t) = p_0 e^{-i\omega t} \sum_{n=-\infty}^{\infty} \left( \sum_{m=-\infty}^{\infty} D_m H^{(1)}_{n-m}(2k_0 d) \right) J_n(k_0 r) e^{in\theta}, \; r' > b \tag{9}$$

Eqn. (5) is nothing but the total acoustic pressure in $a < r' < b$,

$$P^{total}(r',\theta',t)\Big|_{a<r'<b} = p_0 e^{-i\omega t} \sum_{n=-\infty}^{\infty} \left[ E_n H_n^{(1)}(k_0 f(r')) + F_n J_n(k_0 f(r')) \right] e^{in\theta'} \quad (10)$$

The radial acoustic velocity within the shell is,

$$V(r'<b,\theta',t) = \frac{1}{i\rho_{r'}\rho_0\omega}\frac{\partial P^{total}}{\partial r'} = \frac{1}{i\rho_0\omega}\left(\frac{b^2}{r'^2}\right) p_0 k_0 e^{-i\omega t} \sum_{n=-\infty}^{\infty} \left[ E_n H_n^{(1)'}(k_0 f(r')) + F_n J_n'(k_0 f(r')) \right] e^{in\theta'}$$

(11)

where $H_n^{(1)'}(.)$ and $J_n'(.)$ are the derivatives of $H_n^{(1)}(.)$ and $J_n(.)$ respectively with respect to their arguments.

From Eqn. (6) and (8), the total pressure referred to $(r',\theta')$ coordinate system in $r' > b$ is,

$$P^{total}(r',\theta',t)\Big|_{r'>b} = P_{rad}(r',\theta',t) + p_{scat}(r',\theta',t) \quad (12)$$

$$P^{total}(r',\theta',t)\Big|_{r'>b} = p_0 e^{-i\omega t} \sum_{n=-\infty}^{\infty} \left[ \left( \sum_{m=-\infty}^{\infty} C_m H_{m-n}^{(1)}(2k_0 d) \right) J_n(k_0 r') + D_n H_n^{(1)}(k_0 r') \right] e^{in\theta'} \quad (13)$$

The radial acoustic velocity outside the shell in $(r',\theta')$ coordinate system is,

$$V(r'>b,\theta',t) = \frac{1}{i\rho_0\omega}\frac{\partial P^{total}}{\partial r'} = \frac{1}{i\rho_0\omega} p_0 k_0 e^{-i\omega t} \sum_{n=-\infty}^{\infty} \left[ \left( \sum_{m=-\infty}^{\infty} C_m H_{m-n}^{(1)}(2k_0 d) \right) J_n'(k_0 r') + D_n H_n^{(1)'}(k_0 r') \right] e^{in\theta'}$$

(14)

The above total pressure can also be referred to $(r,\theta)$ coordinate system using Eqn. (1) and (9) as,

$$P^{total}(r,\theta,t)\Big|_{r'>b} = P_{rad}(r,\theta,t) + p_{scat}(r,\theta,t) \quad (15)$$

$$P^{total}(r,\theta,t)\Big|_{r'>b} = p_0 e^{-i\omega t} \sum_{n=-\infty}^{\infty} \left[ C_n H_n^{(1)}(k_0 r) + \left( \sum_{m=-\infty}^{\infty} D_m H_{n-m}^{(1)}(2k_0 d) \right) J_n(k_0 r) \right] e^{in\theta} \quad (16)$$

The radial acoustic velocity outside the shell in $(r,\theta)$ coordinate system is,

$$V(r,\theta,t)\Big|_{r'>b} = \frac{1}{i\rho_0\omega}\frac{\partial P^{total}}{\partial r} = \frac{1}{i\rho_0\omega} p_0 k_0 e^{-i\omega t} \sum_{n=-\infty}^{\infty} \left[ C_n H_n^{(1)'}(k_0 r) + \left( \sum_{m=-\infty}^{\infty} D_m H_{n-m}^{(1)}(2k_0 d) \right) J_n'(k_0 r) \right] e^{in\theta}$$

(17)

One can determine the undetermined expansion coefficients $C_n$, $D_n$, $E_n$ and $F_n$ from suitable boundary and continuity conditions.

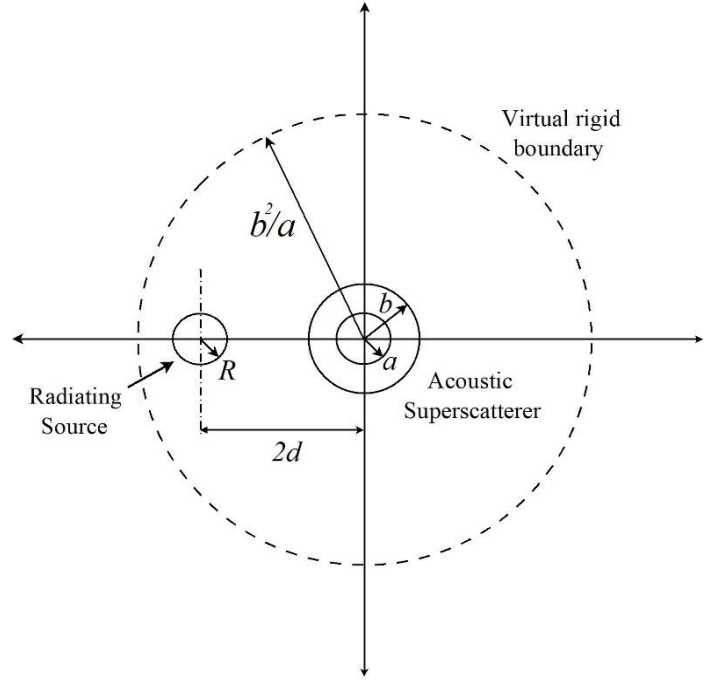

**Fig 3:** The geometrical illustration of the source and superscatterer, showing the virtual rigid boundary, required for applying boundary and continuity conditions

By imposing the continuity of total pressure at $r' = b$ in Eqn. (10) and (13),

$$p_0 e^{-i\omega t} \sum_{n=-\infty}^{\infty} \left[ E_n H_n^{(1)}(k_0 b) + F_n J_n(k_0 b) \right] e^{in\theta'} = p_0 e^{-i\omega t} \sum_{n=-\infty}^{\infty} \left[ \left( \sum_{m=-\infty}^{\infty} C_m H_{m-n}^{(1)}(2k_0 d) \right) J_n(k_0 b) + D_n H_n^{(1)}(k_0 b) \right] e^{in\theta'}$$

(18)

$$E_n H_n^{(1)}(k_0 b) + F_n J_n(k_0 b) = \left( \sum_{m=-\infty}^{\infty} C_m H_{m-n}^{(1)}(2k_0 d) \right) J_n(k_0 b) + D_n H_n^{(1)}(k_0 b) \quad (19)$$

By imposing the continuity of acoustic velocity at $r' = b$ in Eqn. (11) and (14),

$$\sum_{n=-\infty}^{\infty} \left[ E_n H_n^{(1)'}(k_0 b) + F_n J_n'(k_0 b) \right] e^{in\theta'} = \sum_{n=-\infty}^{\infty} \left[ \left( \sum_{m=-\infty}^{\infty} C_m H_{m-n}^{(1)}(2k_0 d) \right) J_n'(k_0 b) + D_n H_n^{(1)'}(k_0 b) \right] e^{in\theta'}$$

(20)

$$E_n H_n^{(1)'}(k_0 b) + F_n J_n'(k_0 b) = \left( \sum_{m=-\infty}^{\infty} C_m H_{m-n}^{(1)}(2k_0 d) \right) J_n'(k_0 b) + D_n H_n^{(1)'}(k_0 b) \tag{21}$$

Applying the Neumann boundary condition corresponding to rigid and impermeable cylinder with radius, $r' = a$, in Eqn. (11),

$$E_n H_n^{(1)'}\left(k_0 \frac{b^2}{a}\right) + F_n J_n'\left(k_0 \frac{b^2}{a}\right) = 0 \tag{22}$$

So we have Eqn. (19), (21) and (22) and four undetermined expansion coefficients $C_n$, $D_n$, $E_n$ and $F_n$. The additional equation required to solve the coefficients is the radiating acoustic pressure from the active source.

The continuity of normal velocity at the surface of the active source requires,

$$Q = e^{-i\omega t} \sum_{n=-\infty}^{\infty} Q_n e^{in\theta} = \left.\frac{\partial P^{total}}{\partial r}\right|_{r=R} \tag{23}$$

where $Q_n$ is the Fourier coefficient representing the radial component of the gradient of modal pressure amplitude of the active cylindrical source [32,33,35]. For monopole vibrational model of active source, $Q_n = p_0 k_0 \delta_{n,0}$, and for dipolar vibrational mode, $Q_n = p_0 k_0 \delta_{n,\pm 1}$, where $\delta_{i,j}$ is the Kronecker delta ($\delta_{i,j} = 1, \forall i = j$ else $0$).

Substituting Eqn. (16) into Eqn. (23) yields,

$$e^{-i\omega t} \sum_{n=-\infty}^{\infty} Q_n e^{in\theta} = p_0 k_0 e^{-i\omega t} \sum_{n=-\infty}^{\infty} \left[ C_n H_n^{(1)'}(k_0 R) + \left( \sum_{m=-\infty}^{\infty} D_m H_{n-m}^{(1)}(2k_0 d) \right) J_n'(k_0 R) \right] e^{in\theta} \tag{24}$$

$$Q_n = p_0 k_0 \left[ C_n H_n^{(1)'}(k_0 R) + \left( \sum_{m=-\infty}^{\infty} D_m H_{n-m}^{(1)}(2k_0 d) \right) J_n'(k_0 R) \right] \tag{25}$$

From Eqn. (19),(21), (22) and (25), the coefficients $C_n$, $D_n$, $E_n$ and $F_n$ can be extracted by suitable truncation of the infinite series followed by matrix inversion method in MATLAB, for a given monopole or dipolar active source.

Our objective is to calculate the far-field pressure amplitude and to demonstrate the effect of superscatterer on it. In far-field region, the asymptotic expressions of cylindrical Bessel and Hankel functions can be substituted in Eqn. (16) resulting in,

$$P^{total}(r,\theta,t) \underset{k_0 r \to \infty}{\approx} p_0 e^{-i\omega t} \sqrt{\frac{2}{\pi k_0 r}} \sum_{n=-\infty}^{\infty} \left[ C_n e^{i\left(k_0 r - \frac{n\pi}{2} - \frac{\pi}{4}\right)} + \left( \sum_{m=-\infty}^{\infty} D_m H_{n-m}^{(1)}(2k_0 d) \right) \cos\left( k_0 r - \frac{n\pi}{2} - \frac{\pi}{4} \right) \right] e^{in\theta}$$

(26)

### 3. Results and Discussion:

#### 3.1. Monopole source

A monopolar oscillating cylinder of radius 2m (R=2m) is placed at a distance of 7m (d=3.5m) from a rigid cylinder of radius 2m (a=2m) coated with a double-negative-metamaterial of inner radius 2m and outer radius 4m (b=4m) in water. The radiating acoustic wave has a pressure amplitude of 1Pa $(p_0 = 1Pa)$, frequency of vibration 1000Hz. The radius of the virtual boundary of the coating is 8m $\left( = b^2/a \right)$ and the source is contained within this region. In the far-field region of 100m ($k_0 r = 419$) away from the source, the total acoustic pressure is monitored and compared with a case wherein no coating on the rigid cylinder is provided. We note excellent suppression of sound radiation in the forward direction with the coating in Fig 4(a). In fig 4(b), the distance between the centres of radiating cylinder and the superscatterer is increased to 9m $\left( > b^2/a \right)$ and the far-field radiation pattern is compared with un-coated case. The result shows an enhanced total pressure in the forward direction for the coated scatterer when compared with an un-coated scatterer.

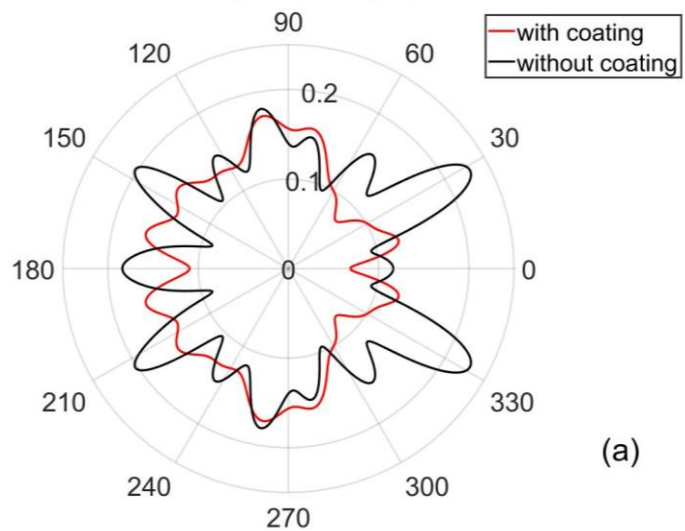

(a)

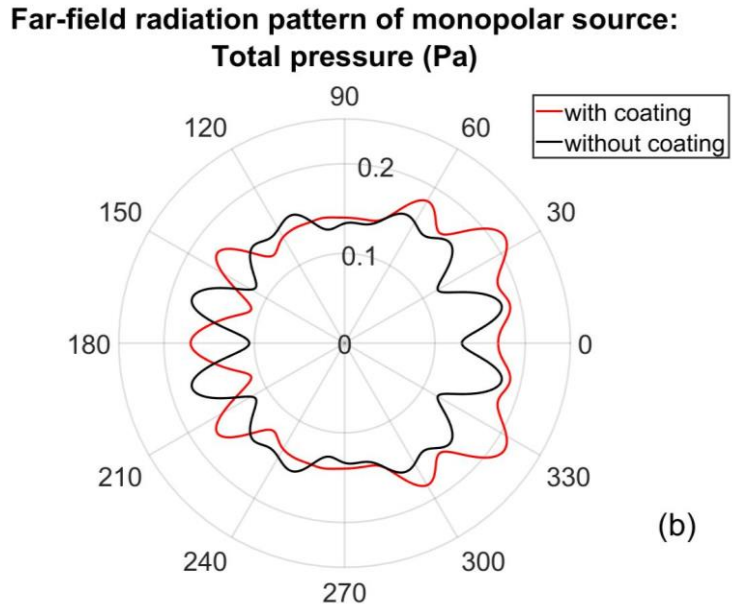

**Fig 4:** Comparison of radiation pattern of total absolute pressure at the far-field ($k_0 r = 419$) for a monopole cylindrical source in the vicinity of a cylindrical scatterer coated with negative metamaterial and without any coating, (a) when axis of cylindrical source is at a distance 7m ($< b^2/a$) from the axis of scatterer, (b) when axis of cylindrical source is at a distance 9m ($> b^2/a$) from the axis of scatterer.

### 3.2. Dipole source

For the same scenario mentioned above the monopole source is replaced with a dipole oscillating cylinder of radius 2m (R=2m) with dipole axis aligned along x-axis and the total acoustic pressure is monitored and compared with a case wherein no coating on the rigid cylinder is provided. The asymptotic behaviour of the total field in forward direction for dipole source is similar to the monopole source as can be seen in Fig. 5.

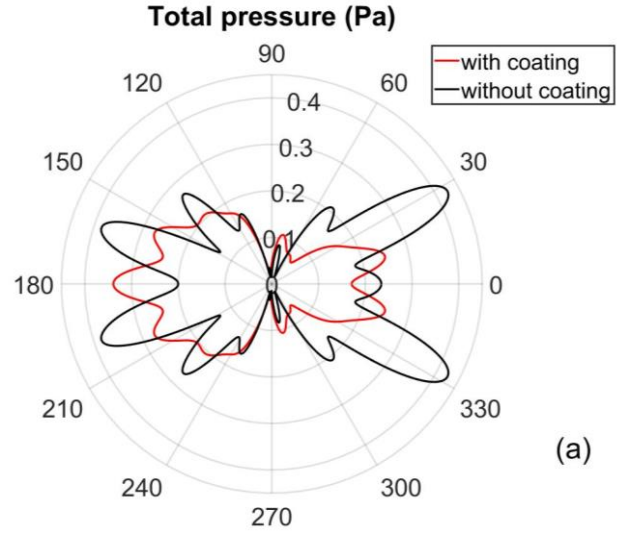

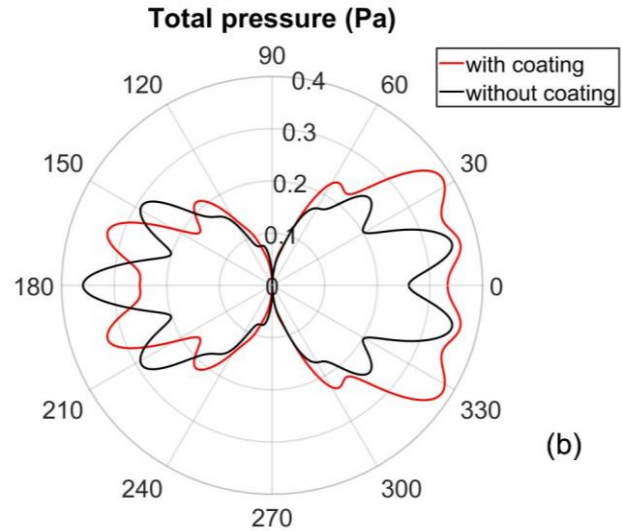

**Fig 5:** Comparison of radiation pattern of total absolute pressure at the far-field ($k_0 r = 419$) for a dipolar cylindrical source in the vicinity of a cylindrical scatterer coated with negative metamaterial and without any coating, (a) when axis of cylindrical source is at a distance 7m ($< {b^2}/{a}$) from the axis of scatterer, (b) when axis of cylindrical source is at a distance 9m ($> {b^2}/{a}$) from the axis of scatterer.

From the results in Fig. 4 and Fig. 5, we can interpret that the total pressure of acoustic wave entering a fixed cylindrical surface at far-field is reduced due to the double-negative metamaterial coatings over the rigid cylindrical scatterer. This can be attributed to the multiple scattering effect between the source and the expanded boundary of the superscatterer as long as the source is contained within it. Once the source crosses this boundary it can view the superscatterer as a magnified object over and above the actual scale.

### 3.3. Extinction cross section:

Any point in space is traversed by two wave systems such as radiated and scattered waves. The flow of energy is a combination of both these waves.

Total time-averaged power of acoustic wave at far-field is,

$$W^{total} = \iint_S \overline{P^{total}\vec{V}} \cdot d\vec{S} \qquad (27)$$

where $P^{total}(r,\theta,t)$ and $V(r,\theta,t)$ are expressed already in Eqn. (16) and (17) respectively.

If we resolve the time-averaged acoustic power of radiating source alone at far-field,

$$W^{rad} = \iint_S \overline{P_{rad}\vec{V_{rad}}} \cdot d\vec{S} \qquad (28)$$

where all quantities in bold are vectors, the integrand with over-bar denotes the time averaged intensity over a period of wave [36] deduced as $\dfrac{P^{total}\vec{V}}{2}$ in Eqn. (27) and $\dfrac{P_{rad}\vec{V_{rad}}}{2}$ in Eqn. (28), and $P_{rad}(r,\theta,t)$ is expressed already in Eqn. (1).

The double integration in Eqn. (27) and (28) is performed over a fixed cylindrical surface of unit length of large radius $r$, and an elemental area vector on this surface is $d\vec{S} = dS\,\hat{n}$ with $\hat{n}$ being the unit outward normal vector on this element. The projections of velocity vectors $(\vec{V}$ and $\vec{V_{rad}})$ along the normal of surface area element, also called radial velocity, as $k_0 r \to \infty$ are [37]: $\vec{V}\cdot\mathbf{n} = \dfrac{P^{total}}{\rho_0 c_0}$, $\vec{V_{rad}}\cdot\mathbf{n} = \dfrac{P_{rad}}{\rho_0 c_0}$.

Then,

$$W^{total} = \iint_S \frac{|P^{total}|^2}{2\rho_0 c_0} dS \qquad (29)$$

$$W^{rad} = \iint_S \frac{|P_{rad}|^2}{2\rho_0 c_0} dS \qquad (30)$$

Substituting Eqn. (26) into Eqn. (29),

$$W^{total} = \iint_S \frac{P^{total} P^{total*}}{2\rho_0 c_0} dS \qquad (31)$$

$$= \frac{p_0^2}{\rho_0 c_0 \pi k_0 r} \sum_{n=-\infty}^{\infty} \sum_{\alpha=-\infty}^{\infty} \int_{\theta=0}^{2\pi} \left[ C_n e^{i\left(k_0 r - \frac{n\pi}{2} - \frac{\pi}{4}\right)} + \left(\sum_{m=-\infty}^{\infty} D_m H_{n-m}^{(1)}(2k_0 d)\right) \cos\left(k_0 r - \frac{n\pi}{2} - \frac{\pi}{4}\right) \right]$$

$$\left[ C_\alpha^* e^{-i\left(k_0 r - \frac{\alpha\pi}{2} - \frac{\pi}{4}\right)} + \left(\sum_{\beta=-\infty}^{\infty} D_\beta^* H_{\alpha-\beta}^{(1)*}(2k_0 d)\right) \cos\left(k_0 r - \frac{\alpha\pi}{2} - \frac{\pi}{4}\right) \right] e^{i(n-\alpha)\theta} r d\theta$$

$$(32)$$

$$= \frac{2p_0^2}{\rho_0 c_0 k_0} \sum_{n=-\infty}^{\infty} \left[ |C_n|^2 + 2\cos\left(k_0 r - \frac{n\pi}{2} - \frac{\pi}{4}\right) \Re\left\{ C_n^* e^{-i\left(k_0 r - \frac{n\pi}{2} - \frac{\pi}{4}\right)} \left(\sum_{m=-\infty}^{\infty} D_m H_{n-m}^{(1)}(2k_0 d)\right) \right\} \right.$$

$$\left. + \left|\sum_{m=-\infty}^{\infty} D_m H_{n-m}^{(1)}(2k_0 d)\right|^2 \cos^2\left(k_0 r - \frac{n\pi}{2} - \frac{\pi}{4}\right) \right] \qquad (33)$$

The terms with asterisk in Eqn. (31-33) denote the complex conjugates.

Similarly, substituting Eqn. (1) into Eqn. (30),

$$W^{rad} = \iint_S \frac{P_{rad} P_{rad}^*}{2\rho_0 c_0} dS \qquad (34)$$

$$= \frac{p_0^2}{\rho_0 c_0 \pi k_0 r} \sum_{n=-\infty}^{\infty} \sum_{\alpha=-\infty}^{\infty} \int_{\theta=0}^{2\pi} C_n C_\alpha^* e^{i\left(k_0 r - \frac{n\pi}{2} - \frac{\pi}{4}\right)} e^{-i\left(k_0 r - \frac{\alpha\pi}{2} - \frac{\pi}{4}\right)} e^{i(n-\alpha)\theta} r d\theta \qquad (35)$$

$$= \frac{2p_0^2}{\rho_0 c_0 k_0} \sum_{n=-\infty}^{\infty} |C_n|^2 \qquad (36)$$

As seen in section 3.1 & 3.2, the total pressure of acoustic wave entering the cylindrical surface of large radius $r$ is reduced due to the double-negative metamaterial coatings over the rigid cylindrical scatterer. The reduction in total acoustic pressure due to interference of the scattered and radiated waves is such as if an area $\sigma_{ext}$ of the radiating object had been covered up [38,39]. $\sigma_{ext}$, called as extinction cross section, is an important measure of the total energy that the scatterer extracts fom the incident wave in the form of radiation or absorption. The mathematical expression for $\sigma_{ext}$ is derived based on conservation of energy which requires zero total energy flux across the closed cylindrical surface of large radius $r$.

$$\sigma_{ext} = \frac{1}{I_0}\left(W^{rad} - W^{total}\right) \tag{37}$$

where $I_0$, the characteristic intensity also known as incident power flux, is denoted by $I_0 = \frac{|p_0|^2}{2\rho c}$.

Substituting Eqn. (33) and (36) into Eqn. (37) yields,

$$\sigma_{ext} = -\frac{4}{k_0}\sum_{n=-\infty}^{\infty}\left[2\cos\left(k_0 r - \frac{n\pi}{2} - \frac{\pi}{4}\right)\Re\left\{C_n^* e^{-i\left(k_0 r - \frac{n\pi}{2} - \frac{\pi}{4}\right)}\left(\sum_{m=-\infty}^{\infty} D_m H_{n-m}^{(1)}(2k_0 d)\right)\right\}\right.$$
$$\left.+\left|\sum_{m=-\infty}^{\infty} D_m H_{n-m}^{(1)}(2k_0 d)\right|^2 \cos^2\left(k_0 r - \frac{n\pi}{2} - \frac{\pi}{4}\right)\right] \tag{38}$$

$\sigma_{ext}$ for both monopole and dipole cylindrical source, parameters as mentioned in section 3.1 and 3.2 respectively, is calculated using Eqn. (38) as a function of the distance between the axis of the source and the scatterer. It can be observed in both Fig 6(a) and 6(b) that when the distance (2d) is 8.4m, the $\sigma_{ext}$ is maximum indicating a strong suppression in total acoustic energy at far-field. This distance, 8.4m, is very close to the virtual rigid boundary, $\frac{b^2}{a} = 8m$, of the superscatterer having complementary metamaterial coating. As the distance increases beyond 8.4m, $\sigma_{ext}$ decreases gradually and approaches zero at very large distances. From the acoustic power aspects, the total acoustic power impinging on a cylinder of fixed radius $r$, for which $k_0 r \to \infty$, is less pronounced for minimum $\sigma_{ext}$. From the observations in Fig. 6(a)&(b) we can infer that the attenuation of total acoustic power at far-field is maximum when the source lies nearer to the virtual rigid boundary of the superscatterer, prominent within the boundary, and reduces gradually when it moves outwards from it.

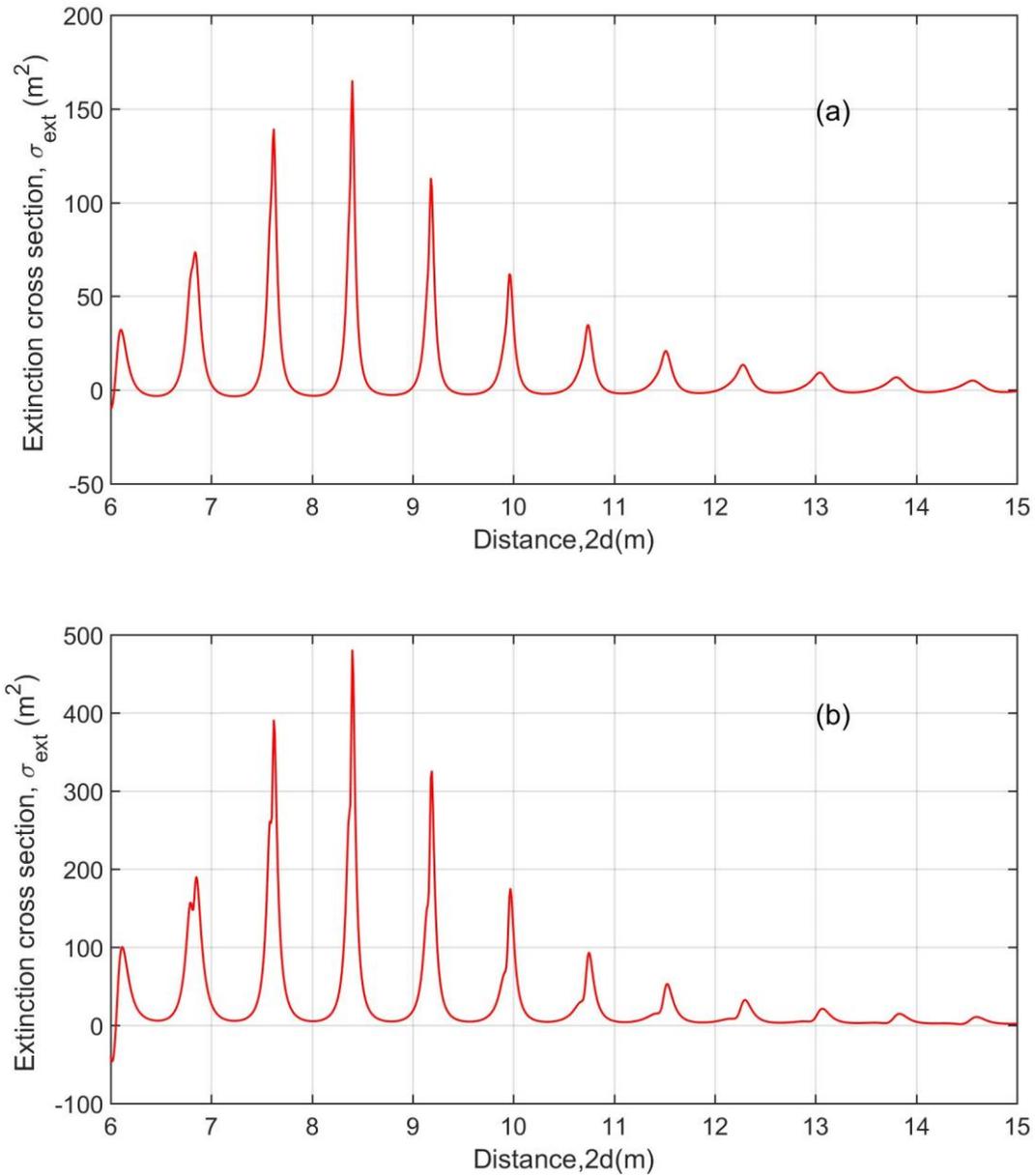

**Fig 6:** Variation of extinction cross section $(\sigma_{ext})$ with respect to the distance between the source and the scatterer at the far-field ($k_0 r = 419$), (a) for a monopole cylindrical source, (b) for a dipolar cylindrical source.

### 3.4 Effect of cross-sectional area of source

The effect of the area of cross section of the cylindrical source is studied in this section. Taking cue from the previous section, the radius of the cylindrical source is varied from 0.1m to 2m for a fixed distance 8m between the source (both monopole and dipole) and the scatterer, for which $\sigma_{ext}$ is maximum.

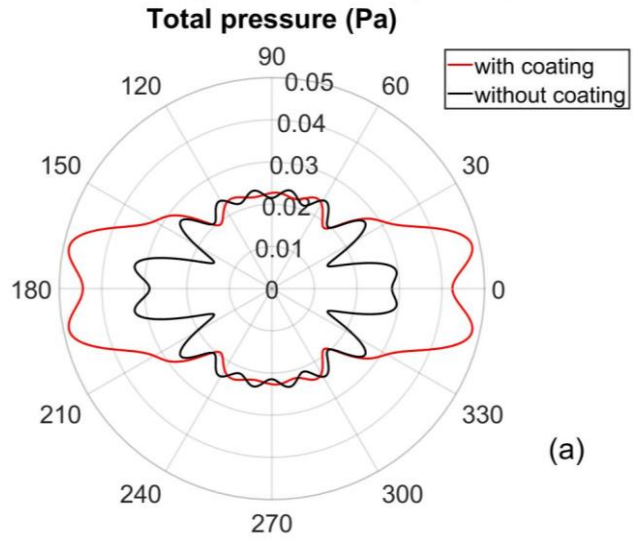

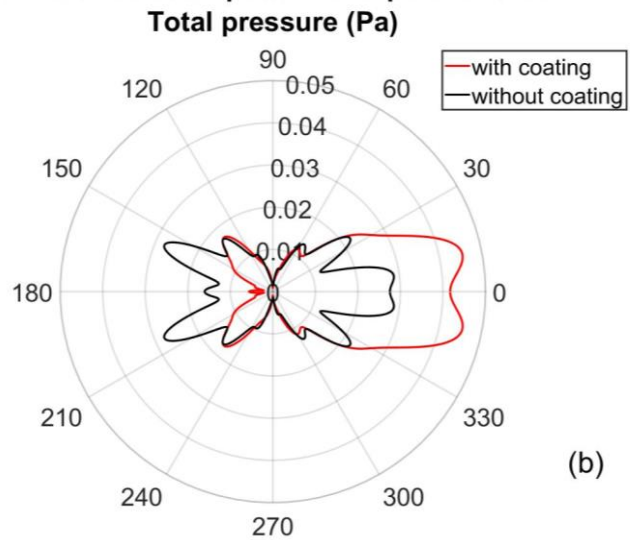

**Fig 7:** Total acoustic pressure at the far-field ($k_0 r = 419$) when the radius of the cylindrical source is 0.1m ($k_0 R = 0.42$) and distance between source and scatterer is 8m, (a) for a monopole cylindrical source, (b) for a dipolar cylindrical source.

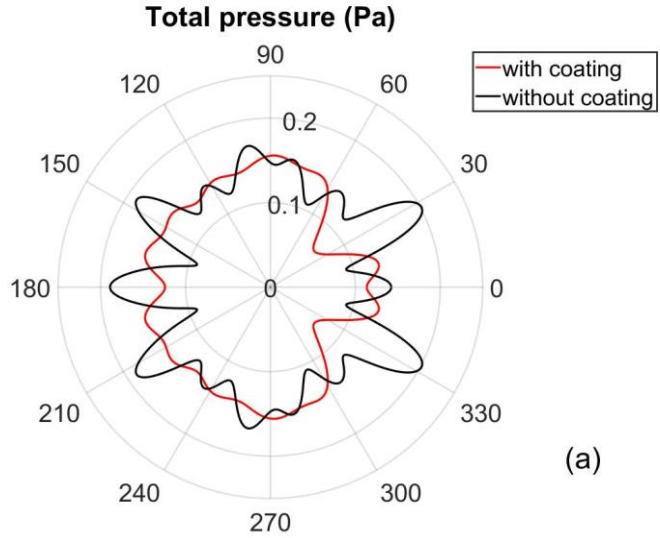

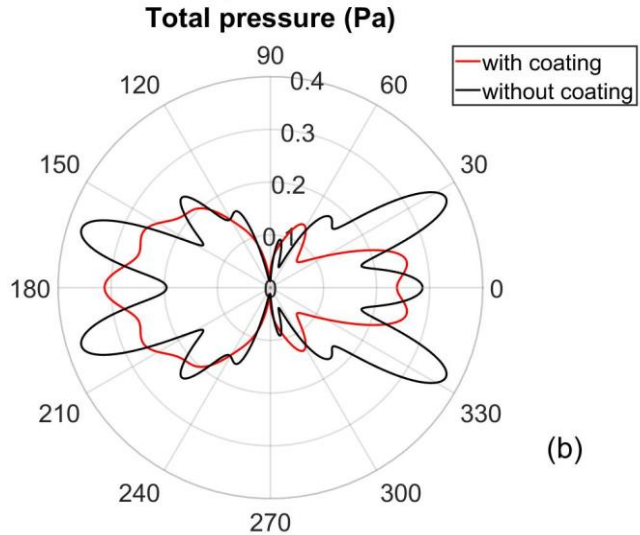

**Fig 8:** Total acoustic pressure at the far-field ($k_0 r = 419$) when the radius of the cylindrical source is 2m ($k_0 R = 8.38$) and distance between source and scatterer is 8m, (a) for a monopole cylindrical source, (b) for a dipolar cylindrical source.

The far-field suppression of acoustic pressure is less pronounced for a Rayleigh active source ($k_0 R \ll 1$) as seen Fig. 7 (a) &(b) when compared to Fig.8 (a)&(b) due to reduction in the multiple scattering effect. This can be further ascertained by calculating the extinction cross section for different cross-sectional areas of the source. The extinction cross section at far-field is plotted with respect to the radius of the cylindrical source in Fig.9, for both monopolar and dipolar source. It can be easily noted that $\sigma_{ext}$ is negligible below R=0.23m (corresponding to $k_0 R \leq 1$) whereas it is the maximum when R=1.62m. This can be attributed to the strong interaction between the reflected wave from the virtual boundary and the surface of the radiating source when R=1.62m.

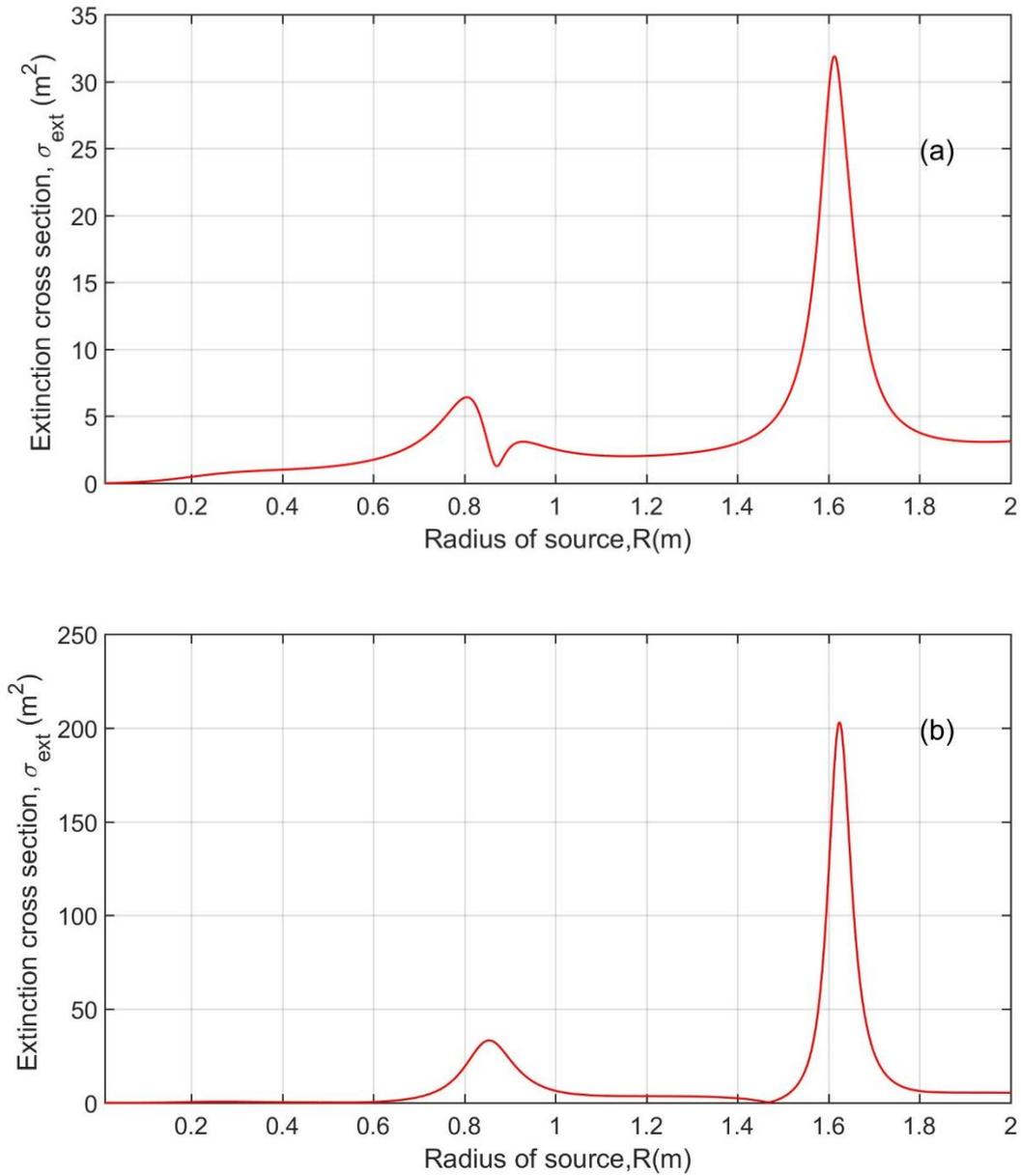

**Fig 9:** Variation of $\sigma_{ext}$ with respect to the radius of the cylindrical active source at the far-field $(k_0 r = 419)$, (a) for a monopole cylindrical source, (b) for a dipolar cylindrical source.

### 3.5 Numerical Analysis

In order to verify the effectiveness of the analytical model of far-field suppression of acoustic field in presence of superscatterer, numerical simulation studies implemented in a commercial package (COMSOL Multiphysics) were performed for the same cases described in section 3.1 and 3.2. The host medium is water and the density tensor in Eqn. (7) is assigned the double-negative metamaterial coating. The relative bulk modulus of the double-negative metamaterial in Eqn. (7) is entered slightly differently due to the following reason. According to Nicorovici *et al.* [29], for a partially-resonant system in an electrostatic field, there is no physical solution to the coated-cylinder problem if the coating dielectric constant $(\varepsilon_s)$ and the exterior medium dielectric constant $(\varepsilon_m)$ satisfy the relation $\varepsilon_s = -\varepsilon_m$ and the radial distance of source from the centre of superscatterer $(r' = Z_0)$ is less than $Z_c \left(= b^3/a^2\right)$. More specifically, the series for the external electric potential $(V_e)$ and the coating potential $(V_s)$ fail to converge in specific regions in space. Additionally, since the image dipole is located in the external medium, creating an unphysical singularity, at which the expansion of $V_e$ diverges; there appears also an image dipole in the coating region, creating an unphysical singularity of $V_s$. To guarantee the existence of a physical solution, a small imaginary part to the dielectric constant of the coating is to be added so that it becomes lossy, i.e.; $\varepsilon_s = (-1+i\delta)\varepsilon_m, |\delta| \ll 1$. The exact duality between 2D Maxwell equations for transverse electric (TE) polarization and the 2D acoustic wave equations in cylindrical coordinates leads to $\lambda^{-1} \leftrightarrow \varepsilon_z$ [40]. It is therefore necessary to add a small imaginary part to the compressibility of the acoustic medium to guarantee the existence of a physical solution to the presented problem. Hence the bulk modulus of the coating, relative to $\lambda_0$, in Eqn. (7) is modified as,

$$\lambda(r') = \left(-\frac{r'^4}{b^4} + i\delta\right), \text{ where } |\delta| \ll 1 \tag{39}$$

Numerical analysis is done by assigning the above complex bulk modulus with $|\delta| = 0.01$ to the double-negative metamaterial and the results are discussed below. For each monopole and dipole radiating source, three cases are simulated and analysed. In the first case, the Sound Pressure Level (SPL) at far-field $(k_0 r = 419)$ when the source is contained within the expanded virtual boundary of the superscatterer as discussed in Section 3.1 & 3.2 is measured numerically. In the second and third case, the far-field SPL in presence of an un-coated rigid scatterer of radius a=2m and a=4m respectively is studied. In order to understand the angular dependence of sound suppression by acoustic superscatterer, polar plots of these 3 cases corresponding to monopole and dipole source are generated and compared in Fig. 10(a) and Fig. 10(b). It can be easily noticed that the forward propagation of the total acoustic field is suppressed in both the cases due to the influence of the superscatterer. Another interesting observation is that the superscatterer (a=2m, b=4m) can scatter more effectively than a rigid scatterer of the same scale (a=4m). From the numerical results

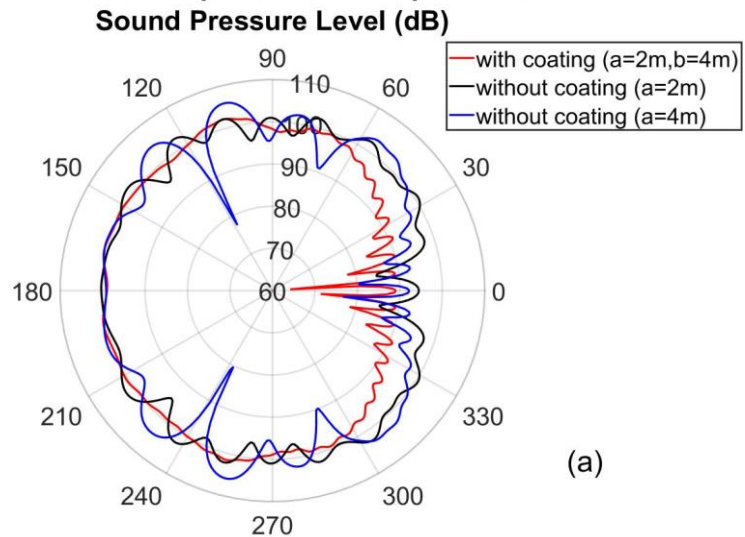

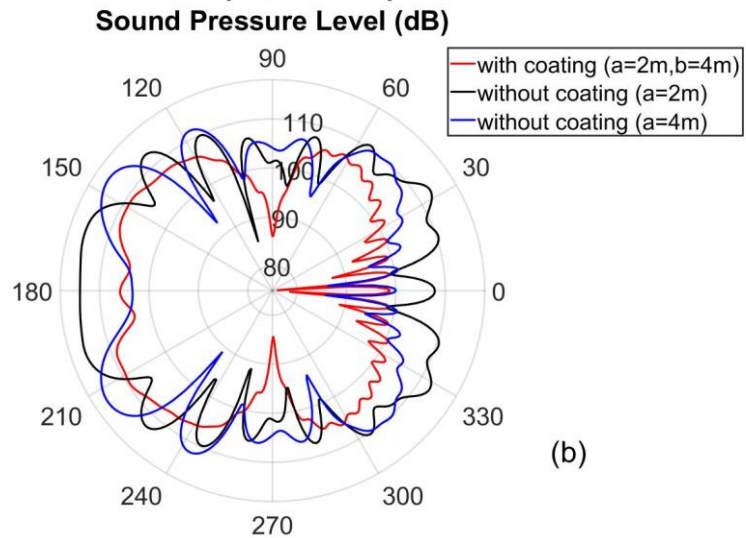

**Fig 10:** Comparison of the far-field SPL obtained numerically for three cases, i.e.; in presence of superscatterer (a=2m, b=4m) when source is contained inside the span of virtual boundary of the superscatterer, in presence of a rigid scatterer (a=2m), and in presence of rigid scatterer (a=4m), corresponding to (a) Monopole radiating source, and (b) dipole radiating source as mentioned in section 3.1 and 3.2 respectively.

**3.6 Applications of the proposed method**

As discussed in the section 1, the proposed method demonstrates passive forward-suppression of acoustic waves emanated from cylindrical sources at far-field distances, useful in many applications, for example, suppression of machinery noise, underwater propeller noise etc. This is a passive method and requires no additional power. Omni-directional damping of sound can be achieved when multiple acoustic superscatterers are placed around the radiating source. By suitably selecting the transformation function the user can easily adjust the virtual boundary of the superscatterer as per the requirement, thereby reducing the unwanted interaction of this boundary with other sources.

**4.    Conclusion**

In summary, we studied the concept of complementary media in acoustics when the source lies within the media and its effect on the sound attenuation analytically. Due to the virtual enhancement of the rigid boundary of superscatterer, acoustic waves radiated from a cylindrical active source undergo multiple reflections between the virtual boundary and the source. It has been shown analytically that the total radiated acoustic pressure is reduced at the far-field in forward-direction for this case as compared to the case when the source lies outside the virtual boundary both for monopolar and dipolar sources. To substantiate these results, the total extinction cross-section of the superscatterer is derived and plotted as a function of the distance between the source and scatterer. The results show that the total extinction cross-section is the highest when the distance coincides with the virtual boundary for both sources. It has also been shown that the radius of the source has a significant effect on the total-radiated far-field pressure. The far-field sound suppression is less pronounced for a Rayleigh active source due to reduction in multiple scattering effect. Finally, the analytical results are validated with numerical results. In total, we have developed an analytical framework for a source-superscatterer schema for passive sound suppression in forward direction.